\newif\ifjournal\journalfalse

\documentclass[aps,prl,reprint,amsmath,amssymb,showpacs]{revtex4-1}

\usepackage{graphicx}
\usepackage{bm}
\renewcommand{\vec}[1]{\boldsymbol{#1}}

\newcommand{\grad}{\nabla}
\renewcommand{\div}{\nabla \cdot}

\def\grl{{Geophys. Res. Lett.}}
\def\jgr{{J. Geophys. Res.}}

\def\prl{{Phys. Rev. Lett.}}
\def\pop{{Phys. Plasmas}}
\def\ssr{{Space Science Reviews}}

\begin{document}

\title{New Measure of the Dissipation Region in Collisionless Magnetic Reconnection}

\author{Seiji Zenitani}
\author{Michael Hesse}
\author{Alex Klimas}
\author{Masha Kuznetsova}
\affiliation{
NASA Goddard Space Flight Center, Greenbelt, Maryland 20771, USA
}

\ifjournal
\date{\today}
\else
\date{Received 7 February 2011; published 11 May 2011}
\fi

\begin{abstract}
A new measure to identify a small-scale dissipation region
in collisionless magnetic reconnection is proposed.
The energy transfer from the electromagnetic field to plasmas in the electron's rest frame
is formulated as a Lorentz-invariant scalar quantity.
The measure is tested by
two-dimensional particle-in-cell simulations in typical configurations:
symmetric and asymmetric reconnection, with and without the guide field. 
The innermost region surrounding the reconnection site is
accurately located in all cases. 
We further discuss implications for nonideal MHD dissipation.
\end{abstract}

\maketitle

Magnetic reconnection \cite{birn07} is a fundamental process in many plasma systems,
ranging from laboratory and solar-terrestrial environments
to extreme astrophysical settings. 
The violation of the ideal condition,
$\vec{E} + \vec{v} \times \vec{B} \ne 0$,
is essential to allow the magnetic flux transport across the reconnection point. 
The critical ``diffusion region'' (DR) where the ideal condition is violated
is of strong interest for understanding the key mechanism of reconnection. 
In collisionless plasmas, since ions decouple first from the magnetic fields, 
it is thought that the DR consists of an ion-scale outer region and
an electron-scale inner region.

In two-dimensional (2D) reconnection problems in the $x$-$z$ plane,
a popular criterion to identify
the innermost ``electron diffusion region'' (EDR) is
the out-of-plane component of the electron nonideal condition, $E^*_y \ne 0$,
where
\begin{eqnarray}
\label{eq:E}
\vec{E}^*= \vec{E} + \vec{v}_e \times \vec{B} = 
- \frac{1}{n_eq} \div \overleftrightarrow{P}_e
- \frac{m_e}{q} \Big( \frac{d\vec{v}_e}{dt} \Big),
\end{eqnarray}
and $\overleftrightarrow{P}_e$ the electron pressure tensor.
In particular, it is known that
the divergence of the pressure tensor
sustains a finite $E_y=E^*_y$ at the reconnection point,
arising from local electron dynamics \cite{hesse99,hesse11}.

Recent large-scale particle-in-cell (PIC) simulations
have shed light on the electron-scale structures
around the reconnection site. 
Earlier investigations \cite{dau06,keizo06} found that 
the EDR identified by $E^*_y \ne 0$ \cite{dau06} or
the out-of-plane electron velocity \cite{keizo06}
extends toward the outflow directions. 
Previous research has suggested that
the EDR has a two-scale substructure:
the inner EDR of $E^*_y>0$ and
the outer EDR of $E^*_y<0$ with a super-Alfv\'{e}nic electron jet
\cite{kari07,shay07}. 
Satellite observations found similar signatures
far downstream of the reconnection site \cite{phan07}. 
The roles of these EDRs are still under debate; 
however,
there is a growing consensus that
only the inner EDR or a similar small-scale region should control
the reconnection rate \cite{shay07,klimas08,hesse08}.
Importantly, it was recently argued that
the outer EDR is non- or only weakly dissipative,
because the super-Alfv\'{e}nic jet and $E^*_y \ne 0$ condition stem from
projections of the diamagnetic electron current in a suitably rotated frame \cite{hesse08}.

Meanwhile, a serious question has been raised by 
numerical investigations on asymmetric reconnection,
whose two inflow regions have different properties such as in reconnection at the magnetopause
\cite{prit09a,mozer10}.
It was found that
various quantities including $\vec{E}^*$ 
fail to locate the reconnection site in asymmetric reconnection,
especially in the presence of an out-of-plane guide field \cite{prit09a}.
Considering the debate on inner or outer EDRs and
the puzzling results in asymmetric reconnection,
it does not seem that
$\vec{E}^*{\ne}0$ is a good identifier of the critical region.

In this Letter, we propose a new measure to identify
a small, physically significant region surrounding the reconnection site.
We construct our measure based on the following three theoretical requirements. 
First, we are guided by the notion that
dissipation should be related to nonideal energy conversion.
Second, we desire a scalar quantity.
If we use a specific component of a vector,
we have to choose an appropriately rotated frame \cite{hesse08}.
Using a scalar quantity instead,
we do not need to find the right rotation
in a complicated magnetic geometry. 
Third, it should be insensitive to the relative motion between
the observer and the reconnection site. 
For example, the reconnection site can retreat away \cite{oka08}, or,
for example, the entire reconnection system may flap
over a satellite due to the magnetospheric motion. 

Our strategy is as follows.
We choose a frame that can be {\itshape uniquely} specified by the observer.
Among several candidates, we choose the rest frame of an electron's bulk motion
because it would be the best one to characterize electron-scale structures. 
Next we consider
the energy transfer from the field to plasmas in this frame,
which is a scalar quantity.
We then expand it with observer-frame quantities.
The obtained measure meets all three requirements.
It is a Lorentz invariant (frame-independent scalar) and
is related to the nonideal energy transfer. 

We follow the spacelike convention ($-$,+,+,+). 
Let us start from the electromagnetic tensor $F^{\mu\nu}$,
\begin{eqnarray}
F^{\mu\nu} = 
\left(
\begin{array}{cccc}
     0 & ~E_x/c & ~E_y/c & ~E_z/c \\
-E_x/c &     0 &   B_z &  -B_y \\
-E_y/c &  -B_z &     0 &   B_x \\
-E_z/c &   B_y &  -B_x &     0
\end{array}
\right)
\end{eqnarray}
Using a 4-velocity $(u^\mu) = \gamma (c, \vec{v})$,
where $\gamma$ is the Lorentz factor $\gamma=[ 1-(v/c)^2 ]^{-1/2}$, 
we obtain a 4-vector of the rest-frame electric field $e^{\mu}$
\cite{anile89},
\begin{eqnarray}
e^{\mu} = F^{\mu\nu}u_{\nu} = \bar{\Lambda}^{\mu}_{\nu} e'^{\nu}.
\end{eqnarray}
Here the prime sign $'$ denotes the properties in the rest frame of an arbitrary motion $u^\mu$,
and $\bar{\Lambda}$ is the inverse Lorentz transformation from the moving frame.
The components of $e^{\mu}$ and $e'^{\mu}$ are given by,
\begin{eqnarray}
\label{eq:e}
(e^\mu) = \Big( \frac{\gamma \vec{v}\cdot\vec{E}}{c},
\gamma ( \vec{E} + \vec{v} \times \vec{B} ) \Big),~
(e'^\mu) = (0, \vec{E'} ).
\end{eqnarray}
We also use the 4-current $(J^\mu) = (\rho_c c, \vec{j})$,
where $\rho_c$ is the charge density.
The current can be split into the conduction current $j^\mu$ and
the convection current, a projection of the motion of the non-neutral frame
\cite{moller72},
\begin{eqnarray}
\label{eq:J}
J^\mu = j^\mu + \rho'_c u^\mu = j^\mu + c^{-2}(-J^{\nu}u_{\nu})u^\mu,
\end{eqnarray}
such that $(j'^\mu) = (0, \vec{j}' )$ is purely spacelike.

Let us define a dissipation measure $D$,
the energy conversion rate in the moving frame.
The contraction of the covariant and contravariant vectors
gives us a Lorentz-invariant scalar,
\begin{eqnarray}
\label{eq:rela1}
D(u) &=& \vec{j}'\cdot\vec{E}'
= j'_{\mu}e'^{\mu}
\equiv j_{\mu}e^{\mu}
\nonumber\\
&=& J_{\mu} e^{\mu} + c^{-2}J^{\alpha}u_{\alpha} (u_{\mu} F^{\mu\nu}u_{\nu})
= J_{\mu} F^{\mu\nu} u_{\nu}
\nonumber \\
&=& \gamma \big[ \vec{j} \cdot ( \vec{E} + \vec{v} \times \vec{B} ) -
\rho_c ( \vec{v} \cdot \vec{E} ) \big]
.
\end{eqnarray}
Choosing the frame of electron bulk motion (the number density's flow),
we obtain the {\it electron-frame dissipation measure},
\begin{eqnarray}
\label{eq:EDR}
D_e =
\gamma_e \big[ \vec{j} \cdot ( \vec{E} + \vec{v}_e \times \vec{B} ) -
\rho_c ( \vec{v}_e \cdot \vec{E} ) \big].
\end{eqnarray}
In the nonrelativistic limit,
one can simplify Eq.~\eqref{eq:EDR} by setting $\gamma_e \rightarrow 1$. 
One can confirm this by multiplying 
$\vec{j}' = qn_i\vec{v}_i' = (\vec{j} - \rho_c \vec{v}_e)$
and $\vec{E}' = \vec{E}^* = (\vec{E} + \vec{v}_e\times\vec{B})$.

In ion-electron plasmas,
since $J^{\mu}=q(n_iu^{\mu}_i-n_eu^{\mu}_e)$, where $n$ is the proper density,
we obtain the following relation between
the electron-frame and ion-frame measures,
\begin{eqnarray}
\label{eq:prop} 
n_e D_e = n_i D_i
.
\end{eqnarray}
Such a symmetric relation is reasonable,
as ions are the current carrier in the electron's frame and vice versa.
If ions consist of multiple species,
$n_e D_e = \sum_{s} Z_s n_s D_s$,
where $s$ denotes ion species and $Z$ is the charge number.


To see how our measure characterizes the reconnection region,
we have carried out 2D nonrelativistic PIC simulations.
The length, time, and velocity are normalized by
the ion inertial length $d_i=c/\omega_{pi}$,
the ion cyclotron frequency $\Omega_{ci}^{-1}$, and
the ion Alfv\'{e}n speed $c_{Ai}$, respectively.
The mass ratio is $m_i/m_e=25$, and
the electron-ion temperature ratio is $T_e/T_i=0.2$.
Periodic ($x$) and conductive wall ($z$) boundaries are used.
Four runs ($1$-$4$) are carried out.
Runs 1 and 2 employ a Harris-like configuration,
$\vec{B}(z)=B_0 \tanh(2z) \vec{\hat{x}}$ and $n(z) = n_{0} [0.2 + \cosh^{-2}(2z)]$.
The domain of $[0,102.4]\times[-25.6, 25.6]$ is resolved by $1600^2$ cells.
$2.6{\times}10^9$ particles are used.
The speed of light is $c=10$.
In run 2, we impose a uniform guide-field $B_y=B_0$. 
Runs 3 and 4 employ asymmetric configuration.
Since no kinetic equilibrium is known, we employ
the following fluid equilibrium proposed by Ref.~\cite{prit08},
$\vec{B}(z) = B_0 [ \frac{1}{2} + \tanh (2z) ]~\vec{\hat{x}}$ and
$n(z) = n_0 \big[ 1 - \frac{1}{3} \tanh(2z) - \frac{1}{3} \tanh^{2}(2z) \big]$.
Across the current sheet,
magnetic fields and the density vary
from $-B_0/2$ and $n_0$ to $3B_0/2$ and $n_0/3$.
The domain of $[0,64]\times[-12.8, 12.8]$ is resolved by
$1000{\times}800$ grid points.
$9{\times}10^8$ particles are used.
The speed of light is $c=20$.
In run 4, a guide-field $B_y=B_0$ is added.
In all runs, reconnection is triggered by a small flux perturbation.

\begin{figure}[thp]
\begin{center}
\ifjournal
\includegraphics[width=0.93\columnwidth]{f1.eps}
\else
\includegraphics[width=0.93\columnwidth]{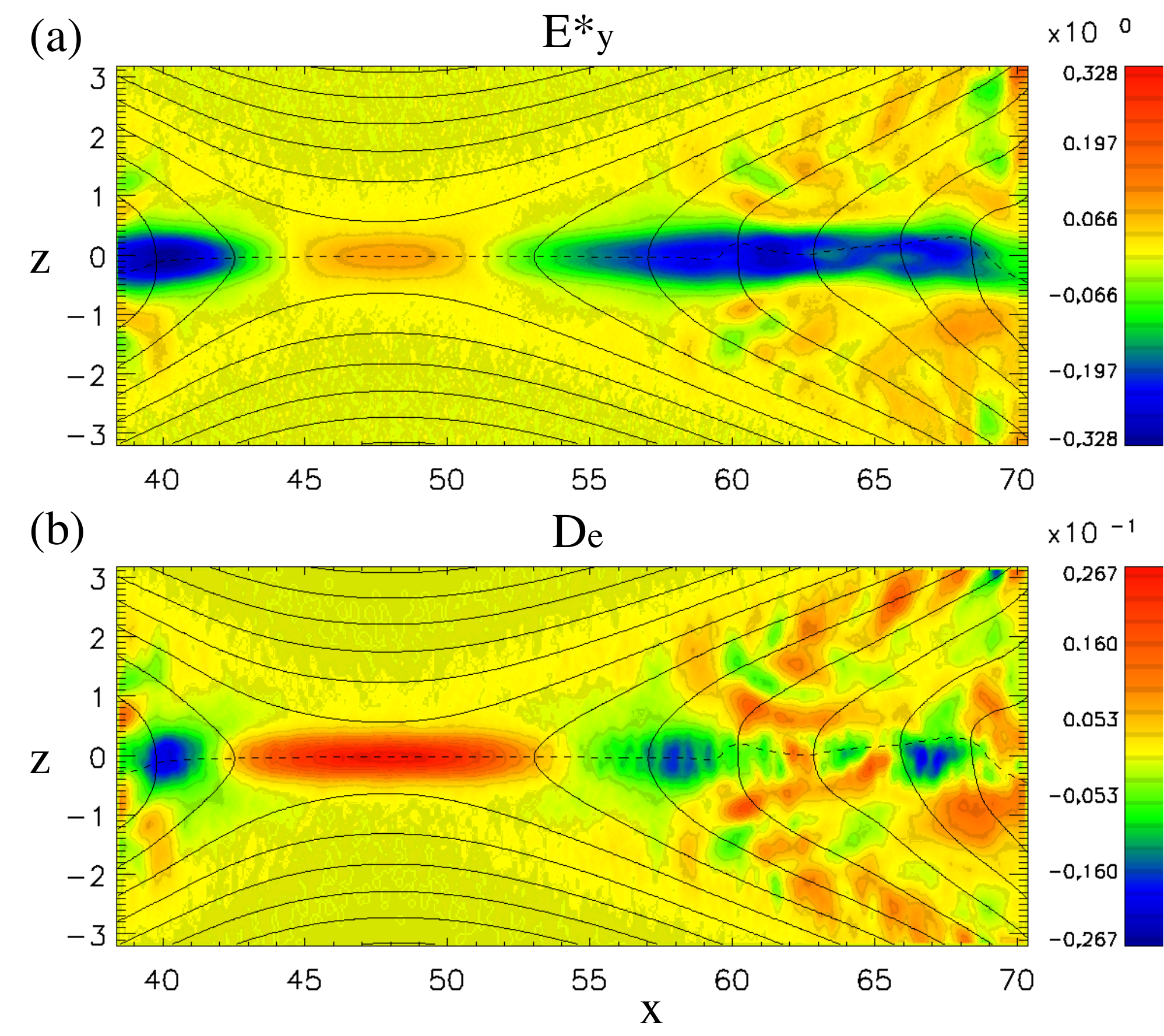}
\fi
\caption{(color online).
\label{fig:panel1}
Snapshots of run 1 at $t=60$, averaged over $\Omega_{ci}^{-1}$.
(a) The nonideal electric field $E^*_y$ and
(b) the electron-frame dissipation measure $D_e$ (Eq.~\eqref{eq:EDR}). 
}
\end{center}
\end{figure}


The panels in Fig. \ref{fig:panel1} present
the popular measure $E^*_y$ and the electron-frame dissipation $D_e$
in run 1 in the well-developed stage.
They are normalized by $c_{Ai}B_0$ and $c_{Ai}B_0j_0$, respectively.
Another option is to employ
the upstream normalization \cite{kari07,shay07} or
its hybrid extension for asymmetric cases \cite{cassak07},
but these are beyond the scope of this paper.
All quantities are averaged over $\Omega_{ci}^{-1}$ to remove noise. 
In Fig. \ref{fig:panel1}(a), one can recognize
a positive $E^*_y$ region near the reconnection site and
a negative $E^*_y$ channel which extends to the outflow direction.
They correspond to the inner and outer EDRs \cite{kari07,shay07}.
On the other hand, Fig. \ref{fig:panel1}b gives a different picture.
There is a positive $D_e$ region near the reconnection site,
indicating that the strong energy transfer occurs there.
Hereafter we call it ``dissipation region'' of $D_e>0$.
At the reconnection point,
a main contributor to $\vec{j}'\cdot\vec{E}'$ is $j_y E'_y$.
In this case, the charge term $\rho_c\vec{v}_e\cdot \vec{E}$ is responsible for
$-25\%$ of the total value,
due to significant charge separation, $|n_i-n_e|/(n_i+n_e)\sim 15\%$.
As we move to the outflow direction,
$j_y E'_y$ is gradually replaced by $j_x E'_x$.
This makes the dissipation region longer than the inner EDR. 
Based on the scale height, 
the aspect ratio of the dissipation region, $9.5:0.72$,
is similar to the universal reconnection rate of $0.1$.
Outside of minor fluctuations,
there are no significant structures in the downstream region ($x>55$).

We have also studied the ion-frame dissipation $D_i$ in this case.
The spatial profile closely resembles to that of $D_e$,
as indicated by Eq.~\eqref{eq:prop}
for such a quasineutral plasma ($n_e \simeq n_i$).


\begin{figure}[thp]
\begin{center}
\ifjournal
\includegraphics[width=0.93\columnwidth]{f2.eps}
\else
\includegraphics[width=0.93\columnwidth]{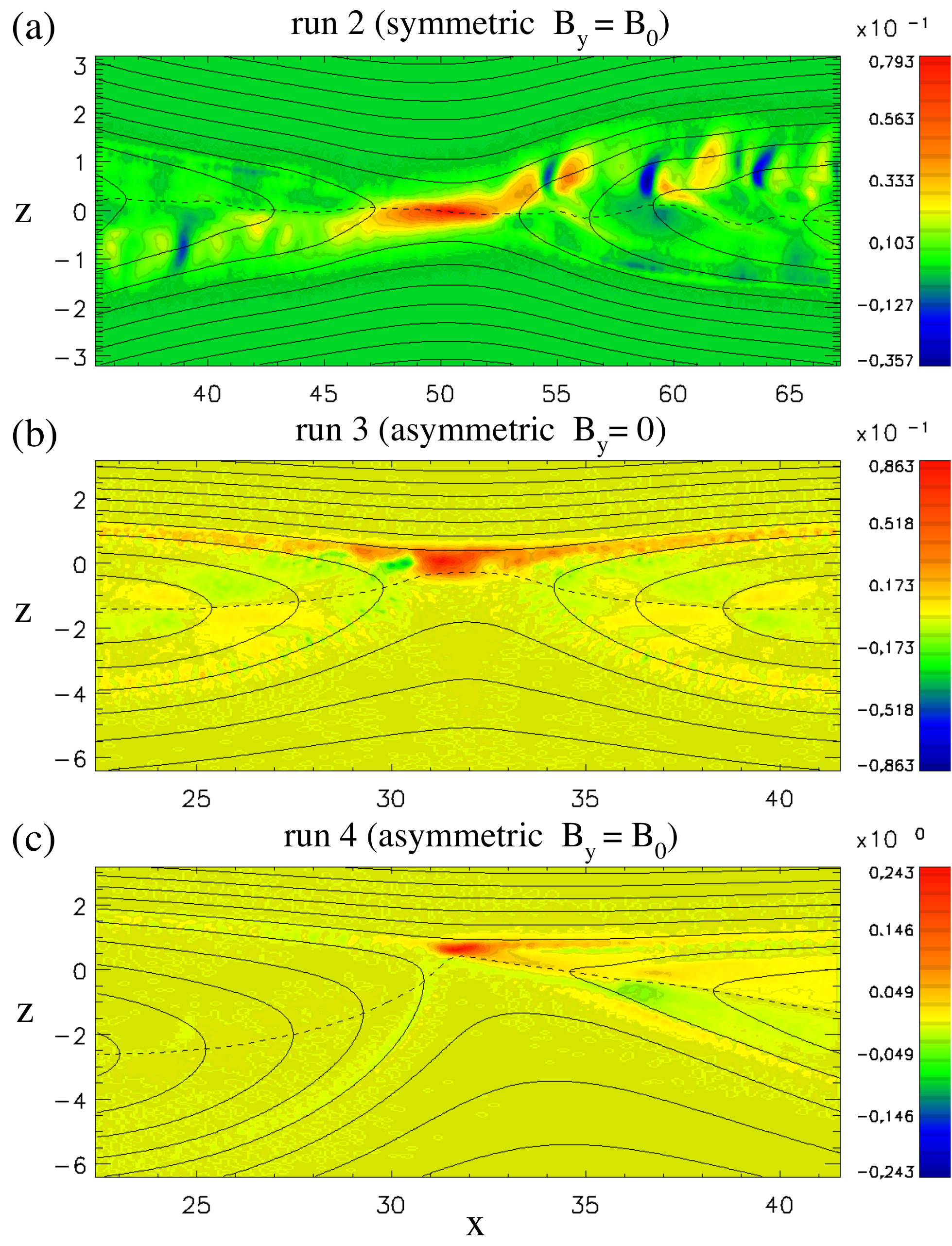}
\fi
\caption{(color online).
\label{fig:panel2}
The dissipation measure $D_e$ (Eq.~\eqref{eq:EDR})
at t=60 in other runs:
(a) run 2,
(b) run 3,
and
(c) run 4.
The dash line show the field reversal line, $B_x=0$.
}
\end{center}
\end{figure}

The panels in Fig. \ref{fig:panel2} show
the dissipation measure $D_e$ in three other runs. 
One can see that it excellently identifies
compact regions about the reconnection sites in all cases.
In the symmetric guide-field run [Fig. 2(a)],
the dissipation region is tilted slightly anticlockwise.
This is associated with the electron cavities
with a parallel electric field along one pairs of separatrices \cite{prit04}. 
In the asymmetric cases, $D_e$ excellently works
even in the most challenging case with a guide-field \cite{prit09a}. 
The field reversal lines ($B_x=0$), shown by the dashed lines,
are located inside our dissipation regions. 
The peak amplitude of $D_e$ is high in the guide field cases,
as the strong electron current is confined.
This deserves further investigation, because
the kinetic dissipation mechanism is different in guide-field cases \cite{hesse11}.
There are charge-separated regions
along the separatrices in the guide-field cases and
on the boundary with the upper inflow regions in the asymmetric cases.
Such charge-separation effects are included in $D_e$
via the last term in Eq.~\eqref{eq:EDR},
which usually improves the identification of the dissipation region.


\begin{figure}[thp]
\begin{center}
\ifjournal
\includegraphics[width=0.93\columnwidth]{f3.eps}
\else
\includegraphics[width=0.93\columnwidth]{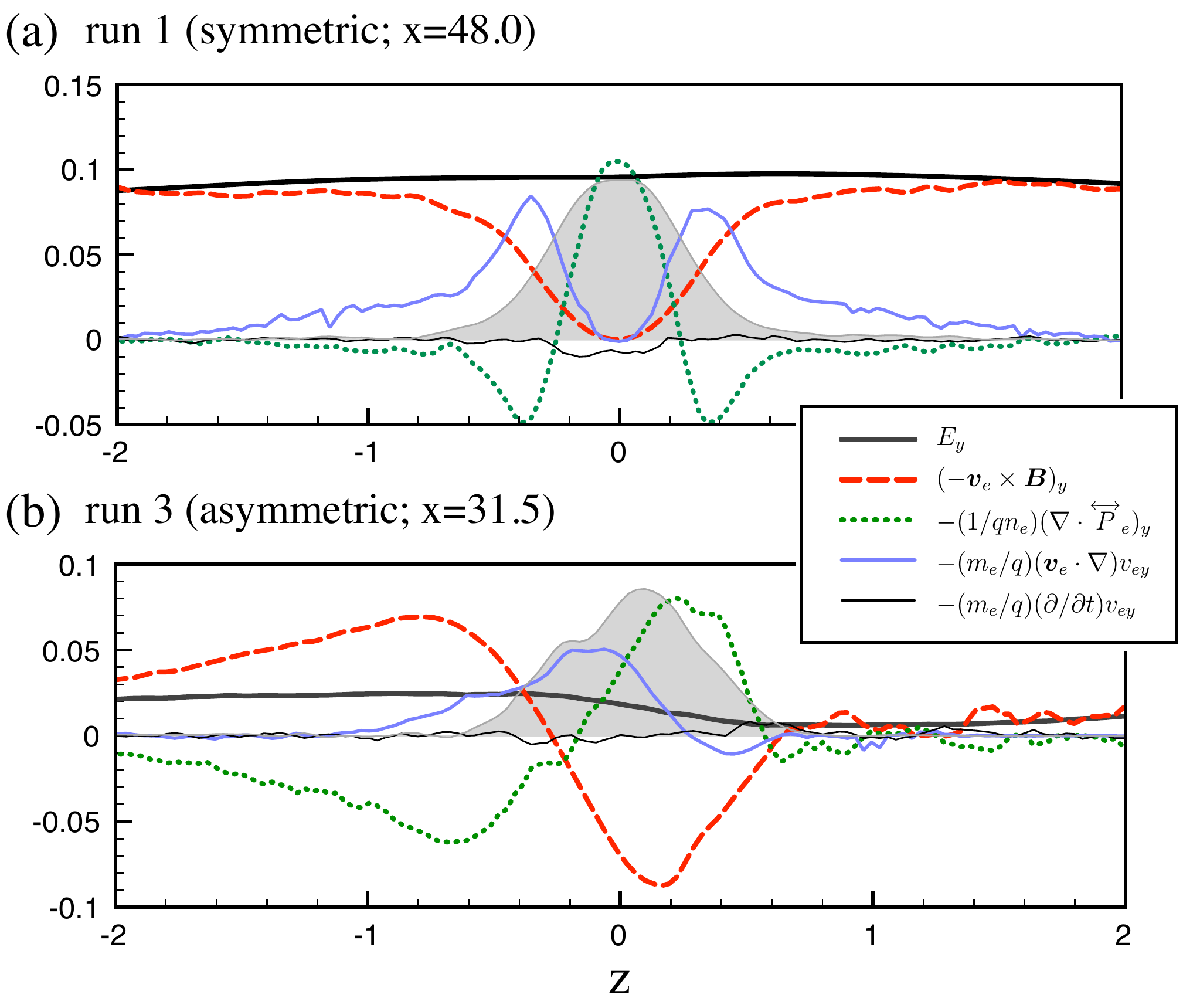}
\fi
\caption{(color online).
\label{fig:1D}
Composition of the reconnection electric field $E_y$ (Eq.~\ref{eq:E}.)
along the inflow line
at (a) $x=48$ in run 1
and
(b) $x=31.5$ in run 3.
The shadow presents a rescaled value of $D_e$
to indicate the dissipation region.
}
\end{center}
\end{figure}

Let us focus on the dissipation region in runs 1 and 3. 
The panels in Fig. \ref{fig:1D} show
the composition of the reconnection electric field $E_y$
along the inflow lines, based on Eq.~\eqref{eq:E}. 
The shadows indicate the rescaled amplitude of the dissipation measure $D_e$.

In the symmetric case [run 1; Fig. 3(a)],
$E_y$ is balanced by the electron pressure tensor term \cite{hesse99} at the center
and the bulk inertial term in surrounding regions.
The dissipation region is located
in a narrow region on local electron inertial scale, $d_e{\sim}0.5$-$6$.
There, the bouncing electrons carry a strong electron current, and
$j'_y$ is intense accordingly.
We also note that
the electron ideal condition is weakly violated on the larger scale of
the local ion inertial length, ${\sim}2$-$3$.
Such an outer structure is related to ion's decoupled motion, and
we will see a clear two-scale structure at sufficiently large times \cite{ishizawa04}.

In asymmetric reconnection, it is known that
the field reversal ($B_x=0$) and the flow stagnation points ($v_{iz},v_{ez}=0$) are
usually not collocated \cite{cassak07,prit08,cassak09}.
In our run [Fig. 3(b)], the electron flow stagnation point is
located on the upper side ($z \simeq 0.63$)
while the field reversal is on the lower side ($z \simeq -0.3$).
Therefore the motional electric field even becomes negative around the dissipation region.
Importantly, $E'_y$ remains positive there and
it resonates with a main current $j_y>0$.
Interestingly, $j_zE'_z$ also contributes to $D_e$ in the lower side ($z\sim-0.3$).
On the other hand, even though $E'_z$ is
an order of magnitude larger than $E'_x$ and $E'_y$
on the upper side ($0<z<0.5$) \cite{prit09a},
it does not contribute to $D_e$ because the vertical current $j_z$ is negligible. 
Physically, such a strong $E'_z$ is overemphasized
by the diamagnetic drift where the density gradient is strong. 
Regarding $E'_y$, we find that
the ideal condition is violated in the lower-side upstream of $z<0$.
The pressure tensor terms, both $\partial_xP_{exy}$ and $\partial_zP_{eyz}$, are not negligible there.
We find that they stem from
the gyrotropic electron pressure tensor
$\overleftrightarrow{P}_e\sim P_{e\perp} \vec{I} + (P_{e\parallel}-P_{e\perp})\vec{B}\vec{B}/B^2$
in the upstream region.
This tells us that these effects are due to the drift motion of gyrating electrons. 
We also notice that the time derivative term ($\partial_t$)
is a key contributor on the upper side ($z\sim 0.6$),
because the reconnection site moves upward very slowly.
Even when the structure is stationary in a frame $d/dt = 0$,
the time derivative $\partial_t = -(\vec{v}_{\rm frame}\cdot \grad)$ is not always zero
in the observer frame.
The fact $\partial_t\ne{0}$ renders the analysis more difficult and
it provides another motivation for using a frame-independent measure.


Let us discuss the relevance for the MHD energy budget.
For simplicity we limit our discussion to the nonrelativistic regime.
Defining an MHD velocity
$\vec{v}_{\rm mhd} =
({m_in_i\vec{v}_i+m_en_e\vec{v}_e})/({m_in_i+m_en_e})$,
we find
\begin{eqnarray}
\label{eq:D}
n_e D_e
&=& n_i D_i
= \frac{m_in_i+m_en_e}{m_i+m_e} D_{\rm mhd}
.
\end{eqnarray}
The total energy transfer 
can be decomposed to
\begin{eqnarray}
\label{eq:jE}
\vec{j} \cdot \vec{E}
= ( \vec{j} \times \vec{B} ) \cdot \vec{v}_{\rm mhd}
+ \rho_c \vec{E} \cdot \vec{v}_{\rm mhd}
+ D_{\rm mhd} 
.
\end{eqnarray}
The first term stands for the work done by the Lorentz force on the MHD fluids.
This operates in the ideal MHD also.
The second term is the work done by the Coulomb force,
also interpreted as the energy transfer by the convection current. 
These two disappear in the MHD frame. 
The last term is responsible for the nonideal energy transfer. 
In a quasineutral plasma $n_i \simeq n_e$,
the positive $D_{\rm mhd}$ ($\simeq D_e$) plays the same role as
an irreversible dissipation given by $\eta j^2>0$ in the resistive MHD.
One can see that positive $D_e$ regions are enhanced and localized
around the reconnection sites in Figs. ~\ref{fig:panel1}(b) and \ref{fig:panel2}.

From the kinetic viewpoint,
gyrating particles undergo various drift motions
such as $\grad{B}$, diamagnetic, and curvature drifts.
When non-$\vec{E}\times\vec{B}$ drifts appear,
the bulk flow no longer comoves with the field lines, and therefore
the ideal condition is no longer a useful concept. 
Drift motions lead to the electromagnetic energy dissipation,
if and only if they involve nonideal energy conversion.
For example, we do not recognize a significant nonideal energy transfer
in the outer EDR in Fig.~\ref{fig:panel1}(b),
where $\vec{E}^*{\ne}0$ 
due to the diamagnetic effect \cite{hesse08}.
On the other hand, around the reconnection sites,
nongyrotropic or field-aligned electrons carry intense currents and then
they enhance the nonideal energy transfer.

In summary,
we have proposed an electron-frame dissipation measure $D_e$ [Eq.~\eqref{eq:EDR}]
to identify a physically relevant, small-scale region surrounding the reconnection point.
We have demonstrated that it works excellently in typical configurations.
Furthermore, we identified its relation to nonideal MHD dissipation,
which is essential to the reconnection problem.

Our finding will benefit several research fields.
One is the satellite observation of reconnection.
NASA is preparing the Magnetospheric Multiscale (MMS) mission
to observe the electron-scale structures in near-Earth reconnection sites.
By using Eq.~\eqref{eq:EDR} with $\gamma_e\rightarrow 1$,
one can identify the dissipation region
regardless of the motion and the orientation of the reconnection site. 
Another example is relativistic astrophysics.
Owning to a growing attention to reconnection,
numerical modeling of relativistic reconnection has been growing in importance \cite{zeni05},
but basic properties are much less known than in the nonrelativistic case.
Our Lorentz-invariant
measure can be readily applied to
the relativistic dissipation region problem \cite{hesse07}.
Further numerical work is desirable to further test our measure
in three dimensions for these systems.

\begin{acknowledgments}
One of the authors (S.Z.) acknowledges support from
JSPS Postdoctoral Fellowships for Research Abroad. This work was supported by NASA's MMS mission.
\end{acknowledgments}


\begin{thebibliography}{}
\bibitem[Birn \& Priest(2007)]{birn07}
J. Birn, E. R. Priest, ``Reconnection of Magnetic Fields: Magnetohydrodynamics and Collisionless Theory and Observations,'' Cambridge University Press (2007)
\bibitem[Hesse et al.(1999)]{hesse99}
M. Hesse {\itshape et al.}, \pop, {\bf 6}, 1781 (1999)
\bibitem[Hesse et al.(2011)]{hesse11}
M. Hesse {\itshape et al.}, \ssr, doi:10.1007/s11214-010-9740-1 (2011)
\bibitem[Daughton et al.(2006)]{dau06}
W. Daughton {\itshape et al.}, \pop, {\bf 13}, 072101 (2006)
\bibitem[Fujimoto(2006)]{keizo06} K. Fujimoto, \pop, {\bf 13}, 072904 (2006)
\bibitem[Karimabadi et al.(2007)]{kari07}
H. Karimabadi {\itshape et al.}, \grl, {\bf 34}, L13104 (2007)
\bibitem[Shay et al.(2007)]{shay07}
M. A. Shay {\itshape et al.}, \prl, {\bf 99}, 155002 (2007)
\bibitem[Phan et al.(2007)]{phan07}
T. D. Phan {\itshape et al.}, \prl, {\bf 99}, 255002 (2007) 
\bibitem[Klimas et al.(2008)]{klimas08}
A. Klimas {\it et al.}, \pop, {\bf 15}, 082102 (2008)
\bibitem[Hesse et al.(2008)]{hesse08}
M. Hesse {\itshape et al.}, \pop, {\bf 15}, 112102 (2008)
\bibitem[Pritchett \& Mozer(2009a)]{prit09a}
P. L. Pritchett and F. S. Mozer, \pop, {\bf 16}, 080702 (2009)
\bibitem[Mozer \& Pritchett(2010)]{mozer10}
F. S. Mozer and P. L. Pritchett, \ssr, doi:10.1007/s11214-010-9681-8 (2010)
\bibitem[Oka et al.(2008)]{oka08} M. Oka {\it et al.}, \prl, {\bf 101}, 205004 (2008)

\bibitem[Anile(1989)]{anile89}
A. M. Anile, ``{Relativistic fluids and magneto-fluids},'' {Cambridge Univ. Press} (1989)
\bibitem[M{\o}ller(1972)]{moller72}
C. M{\o}ller, ``The theory of relativity,'' Oxford: Clarendon Press (1972)

\bibitem[Pritchett(2008)]{prit08}
P. L. Pritchett, \jgr, {\bf 113}, A06210 (2008)
\bibitem[Cassak \& Shay(2007)]{cassak07} P. A. Cassak and M. A. Shay, \pop, {\bf 14}, 102114 (2007)
\bibitem[Pritchett \& Coroniti(2004)]{prit04}
P. L. Pritchett and F. V. Coroniti, \jgr, {\bf 109}, A01220 (2004)
\bibitem[Ishizawa et al.(2004)]{ishizawa04}
A., Ishizawa {\it et al.}, \pop, {\bf 11}, 3579 (2004)
\bibitem[Cassak \& Shay(2009)]{cassak09} P. A. Cassak and M. A. Shay, \pop, {\bf 16}, 055704 (2009)

\bibitem[Zenitani \& Hoshino(2005)]{zeni05}
S. Zenitani and M. Hoshino, \prl, {\bf 95}, 095001 (2005)
\bibitem[Hesse \& Zenitani(2007)]{hesse07}
M. Hesse and S. Zenitani, \pop, {\bf 14}, 112102 (2007)

\end{thebibliography}
\end{document}